# Non-volatile programmable silicon photonics using an ultralow loss $Sb_2Se_3$ phase change material

Matthew Delaney[1,2], Ioannis Zeimpekis[1], Han Du[1], Xingzhao Yan[1], Mehdi Banakar[1], David J. Thomson[1], Daniel W. Hewak[1], Otto L. Muskens[2]*

**Affiliations**

[1]Zepler Institute, Faculty of Engineering and Physical Sciences, University of Southampton, SO17 1BJ, Southampton, UK.

[2] Physics and Astronomy, Faculty of Engineering and Physical Sciences, University of Southampton, SO17 1BJ, Southampton, UK.

*Correspondence to: O.Muskens@soton.ac.uk.

**Abstract**

Adaptable, reconfigurable and programmable are key functionalities for the next generation of silicon-based photonic processors, neural and quantum networks. Phase change technology offers proven non-volatile electronic programmability, however the materials used to date have shown prohibitively high optical losses which are incompatible with integrated photonic platforms. Here, we demonstrate the capability of the previously unexplored material $Sb_2Se_3$ for ultralow-loss programmable silicon photonics. The favorable combination of large refractive index contrast and ultralow losses seen in $Sb_2Se_3$ facilitates an unprecedented optical phase control exceeding $10\pi$ radians in a Mach-Zehnder interferometer. To demonstrate full control over the flow of light, we introduce nanophotonic digital patterning as a conceptually new approach at a footprint orders of magnitude smaller than state of the art interferometer meshes. Our approach enables a wealth of possibilities in high-density reconfiguration of optical functionalities on silicon chip.

**Main Text**

The birth of coherent nanophotonic processors, photonic tensor cores, quantum computing and neuromorphic networks signifies a large paradigm shift towards emerging optical information platforms *(1-6)*. Post-fabrication programming of devices is one of the most desirable functionalities for agile reconfigurable photonic functionalities *(7,8)*. Despite the great success of implementations based on cascaded interferometer meshes *(2, 9-12)*, there are inherent strong limitations in footprint, scalability and volatility. Therefore, the development of new concepts and technologies is of extreme interest.

The fundamental benefits of using non-volatile phase change materials (PCMs) in reconfigurable photonics have resulted in their extensive exploration for photonic modulation and resonance tuning, with $Ge_2Sb_2Te_5$ (GST) *(13-15)* and, more recently, $Ge_2Sb_2Se_4Te_1$ (GSST) *(16)*, being the most considered materials. The multiple non-volatile states these materials offer provide unparalleled energy per bit operation in a highly stable platform. Indeed commercial GST phase change memory has been shown to be stable over $10^{12}$ write cycles. Both materials operate based on a large change in complex refractive index $ñ=n+ik$ between their crystalline and amorphous phases. Despite improvements in materials design, the absorption losses in one or both phases of current PCMs prevent optical phase control independent of changes in the amplitude of propagated light in the telecommunication band, severely limiting phase modulation schemes.

In several recent studies, the antimony-based chalcogenides $Sb_2S_3$ *(17)* and $Sb_2Se_3$ *(18)* have been identified as a family of highly promising ultralow loss PCMs for photonics applications. The materials exhibit no intrinsic absorption losses ($k < 10^{-5}$) in either phase over the telecommunications transmission band and show a low switching temperature around 200 °C whilst remaining non-volatile at operating temperatures. Furthermore, the proximity of their refractive index to that of silicon allows for straightforward direct integration of PCM patches onto standard silicon-on-insulator (SOI) integrated photonics platforms with excellent mode matching to the SOI waveguide.

Here, we demonstrate the exceptional capabilities of the PCM $Sb_2Se_3$ for use in ultralow loss optical phase control of photonic integrated circuits. To achieve this, we make use of 23 nm thin patches of $Sb_2Se_3$ deposited on top of a 220 nm SOI rib waveguide, where the thickness of materials is chosen to maintain a single mode of propagation. An optical phase shift of the propagating wave is induced by changing the material of the PCM from a crystalline to an amorphous state. In our studies, following the example of optical storage media, we start from a fully crystallized PCM as this provides a uniform background, fast writing speeds and stable amorphous written areas. Switching of individual pixels in the PCM is achieved using tightly focused optical pulses from a diode laser operating at 638 nm wavelength (see Methods). From mode calculations we found that an amorphous ($n = 3.29$) single pixel in an otherwise crystalline ($n = 4.05$) $Sb_2Se_3$ layer changes the effective refractive index of the waveguide, $n_{eff}$, by an amount $\Delta n_{eff} = -0.072$.

By selective switching of the PCM in one arm of an asymmetric Mach Zehnder Interferometer (MZI), the optical phase shift is translated into a change of the transmission function of the device. A first demonstration is shown in Fig. 1A in which we achieve phase tuning of an MZI with a 125 µm long patch of $Sb_2Se_3$. The figure shows a succession of vertically offset recorded spectra (grey) from the initial state (red) to the fully switched state (black), followed by a reset to the final state (blue). Spectra were taken every 25 pixels, where each 750nm diameter pixel is spaced by 1000 nm along the MZI. A detailed view shown at the bottom of Fig. 1a shows the initial, set and reset states, with the insertion loss for each spectrum normalized to the initial spectrum. The insertion loss remains unchanged during this process highlighting the lack of any losses induced by the amorphization or recrystallization. A reversible $2\pi$ phase shift was obtained as a result of 100 separate crystallization/amorphization pulses. Regardless of the small imperfections introduced by the motorised stage (backlash) during recrystallization, the full $2\pi$ range was set and reset with a 0.1 nm offset which results in an accumulated error within 2.5% of the full range. More importantly, this measurement demonstrated that a 750 nm pixel size results in a resolution of $0.02\pi$ shift allowing for very fine phase control.

As the induced optical phase change scales linearly with amorphized distance and given the ultralow loss of the material, very large phase tuning ranges can be readily achieved by employing longer PCM structures. Figure 1B shows a $10\pi$ phase shift obtained by switching a 250 µm length of PCM patch, with no measurable change to insertion loss or modulation depth. The effect of the PCM on insertion loss and extinction of the MZIs is explored in more detail in the Supporting Information Fig. S2 which shows that for PCM patch lengths of up to 250 µm, no additional rebalancing of the MZI is required to compensate for losses in the PCM. Apart from using the focused laser spot size for calibration of the switched pixel, the induced optical phase can be fine-tuned by small changes in the laser power. In fact the phase shift was tuned to $0.04\pi$ per pixel in the experiment of Fig. 1B by a small increase in optical peak power of the diode write laser.

To complement the broadband measurements, we performed additional narrowband measurements on a MZI with a shorter, 50 µm patch length as shown in Fig. 1C. Starting at the wavelength of 1553 nm, we tuned the MZI from the maximum transmission (set to 0 dB) to the minimum (-17 dB) and subsequently reset part of this device by recrystallization of a 20 µm length of the PCM patch (red curve). The curve shows the individual levels in transmission induced by switching subsequent pixels in steps of 500 nm along the PCM. This result highlights the ability to define stable multilevel switching processes for highly dense and accurate modulation with very fine quantization.

Next, the endurance of the phase change was probed by repeatedly switching a single pixel of the MZI. We used a pixel at the -13.5 dB point of the MZI response curve on the steepest part of the slope where the sensitivity to individual pixels is high. By repeatedly cycling between the amorphous and crystalline phases in a single spot, the transmission modulation gives a good indication of the lifetime of the phase change. Figure 1D shows the transmission in each phase for the first 600 switches, with red dots corresponding to the crystalline and blue to the amorphous material phase. A long-term drift due to environmental fluctuations of the device transmission was observed and we used a 50-point moving average to normalize the drift as shown in the bottom panel of Fig. 1D. The first 350 full set-reset cycles resulted in a constant transmission change, with a small increase seen towards the 100th cycle. This increase was observed in most tested devices and is attributed to the settling of the switching dynamics after the first few burn-in cycles of the PCM resulting in a slightly increased responsivity. A reduction of the switching contrast below 50% of the initial value is seen after around 500 cycles. Another example of an endurance experiment is shown in the Supporting Information Fig. S4. We note that the endurance seen in our work is a factor 10 lower than that observed for $Sb_2Se_3$ on planar films *(18)* which is attributed to the somewhat more challenging thermal environment of the PCM on the waveguide and may require further improvements in thermal design.

Having successfully demonstrated than an optical phase shift can be induced by switching thin $Sb_2Se_3$ patch on top of a MZI, we moved on to demonstrate a new approach for a programmable optical router based on a multimode interference (MMI) patch. MMIs are of interest for their small footprint as well as for their significantly reduced sensitivity to the environment when compared to MZIs. The device principle is based on the concept of wavefront shaping where a distribution of weak perturbation pixels is used to effectively steer the light with very low loss *(19)*. The perturbation induced by switching of the PCM layer is sufficient to apply this device concept in a practical scalable device configuration. Furthermore, a high level of control is achieved allowing us to independently set and reset pixels spaced at 800 nm pitch in complex patterns.

A 1 x 2 MMI device of $33 \times 6$ µm², covered with a 23 nm crystalline $Sb_2Se_3$ film, was simulated using a 2D FDTD approach, and a pixel perturbation pattern was optimized using an iterative scheme *(19)*. By switching each pixel on the MMI from the input to the output ports, we maximized the transmission from one of the outputs. Repeating this optimization two times resulted in a pattern with high single-port transmission. Figure 2A and B show the calculated intensity in the device before and after patterning.

The designed pattern was subsequently written onto an experimental device. Figure 2C,D show the fabricated MMI with the same dimensions and cladding as the simulated ones, before phase change (C) and after the perturbation pattern was written (D). A residual structure can be seen in the MMI in the unperturbed state (Fig. 2C), which is caused by the crystalline domains of the Sb2Se3 in the crystalline (unswitched) state. In Fig. 2D, a very good agreement is seen in the pattern registration, with only a small ~0.5° tilt present in the experimental pattern due to limitations in the alignment of the setup. Each pixel was written sequentially in columns and rows from the input to the output.

Figure 2E,F show the simulated (dashed curves) and experimental (solid lines) transmission in the top and bottom outputs for both the set (amorphization) and reset (crystallization) steps. Experimentally a 92%:8% splitting ratio was found between the top and bottom outputs respectively for the full pattern. Running the final pattern from the 2D optimization sweep in a full 3D simulation for the fully patterned MMI provides a 97%:3% splitting ratio as shown by the dots in Fig. 2E. The 3D simulation results indicate that optimization using a 2D effective-index model provides a good approximation despite the complex 3D geometry of the perturbation positioned on top of the waveguide. The 5% difference between experimental and simulated results is attributed to experimental limitations in pattern registration onto the device which are as yet not fully understood. The device reset was done pixel-wise in the same way as the write sequence, i.e. from start to end of the pattern, therefore the curves in Fig. 2E and F are not the inverse of each other. Also we point out that the reset was not perfect, this is likely due to some small changes to the film stoichiometry at the edges, where the thermal properties of the MMI are different due to the surrounding insulating $ZnS:SiO_2$ capping layer changing the quenching dynamics.

Other examples of writing pixel patterns are presented in the Supporting Information Fig. S7 and S8, where we used an alternative scheme of iterative optimization of the actual device. Rather than pre-calculating the pixel pattern and writing it to the MMI, here we switched individual pixels and tested their effectiveness on the target transmission. If a pixel contributed positively to the target, its switched state was preserved, otherwise it was erased. Whereas a single pass optimization already resulted in a significant power splitting between the two ports, additional passes are shown to further improve the target transmission function.

Both pre-calculated and iterative optimization patterns result in a target performance exceeding 8 dB extinction between the transmission from the two ports. Importantly, the impact of the patterning on the total device throughput is not strongly affected with up to 0.5 dB additional loss for all devices under study. In some cases an improvement in device transmission was seen, as shown for example in Fig. S8. Such a small improvement in overall device throughput can be attributed to the effect of iterative wavefront shaping which compensates for some of the imperfections of the original device, as seen in earlier studies *(19)*.

To show that the pixel pattern writing approach is both reconfigurable and free-form, the same pattern was written into an MMI multiple times, with a full reset between each pattern for both the simulated pattern and the mirror image of the pattern in the vertical direction which, due to symmetry, guides light towards the bottom output. Figure 3A shows the MMI before (1) and after (2) the first write cycle. To improve the visibility of any changes with respect to the original

MMI patterns, images of all set and reset states were processed by differencing with the original MMI as shown in Fig. 3B. A good repeatability was shown between each pattern, with the reset state showing no memory of the patterns written. After several switching cycles, incremental damage to the $Sb_2Se_3$ film occurred at the edges of the MMI. The damage appears from the corners of the device where the lower local thermal conductivity of the surrounding $ZnS:SiO_2$ capping layer and the abrupt interfaces result in a higher maximum temperature, making the $Sb_2Se_3$ film more prone to delamination or void formation. These effects can be compensated by a variety of methods such as reducing the pulse power for these areas or by reducing the actively written area of the MMI. Figure 3B shows the transmission of both outputs for all the perturbations when writing and resetting the patterns shown in Fig. 3A. A repeatable transmission change was seen for both orientations, which was not strongly affected by the increasing presence of the damaged areas (see Fig. 3A). The splitting ratio was higher for the inverse pattern, suggesting either a small misalignment between the pixel pattern and the MMI, or a non-symmetrical design due to fabrication tolerances in the $Sb_2Se_3$ film.

The results presented here unequivocally show the capabilities of the new phase change technology in providing low-loss programmable optical phase control. This approach is expected to open many new applications in post-fabrication device tuning, programmable weight banks, unitary matrix operations and ultimately all-optical field-programmable arrays *(20)*. We envision that important aspects enabling the mass commercial use of this technology such as the reproducibility, switching endurance, optical and electrical switching will be addressed in follow-on studies, as has been the case with decades of work in optimizing materials like GST.

In conclusion, we have demonstrated the first low loss reconfigurable approach for optical routing in an integrated silicon photonics device. The new phase change material $Sb_2Se_3$ allows the decoupling of optical phase control from amplitude modulation seen in the conventional PCMs. The advances in device footprint and energy consumption of this approach compared to conventional cascaded switch fabrics could enable a range of complex photonic circuits needed for applications such as on-chip LiDAR, photonic quantum technology, AI hardware, or optical tensorcores of the future, whilst providinga powerful post-fabrication programming technique for high-volume PIC ecosystems. Our demonstrated technique provides a general approach that could be easily extended to larger devices and could ultimately achieve a platform for a universal optical chip technology.

**Acknowledgments:** The authors thank N. Dinsdale and P. Wiecha for discussions. **Funding:** This work was supported financially by EPSRC through grant EP/M015130/1. OLM acknowledges support through EPSRC fellowship EP/J016918/1. Silicon photonic waveguides were manufactured through the UK Cornerstone open access Silicon Photonics rapid prototyping foundry through the EPSRC grant EP/L021129/1. D. J. Thomson acknowledges funding from the Royal Society for his University Research Fellowship; **Author contributions:** OLM, IZ, and DH conceived the research and supervised MD. MD and IZ fabricated the phase change materials. MD designed the experimental SOI waveguides with support from DJT. HD, XY, MB and DJT fabricated the etched silicon photonic waveguide devices. MD performed numerical modelling, MD and OLM conducted the optical phase change experiments. MD performed broadband swept source characterizations with support from DJT. MD, IZ, DH, and OLM wrote the manuscript with input from all authors; **Competing interests:** Authors declare no competing interests. and **Data and materials availability:** All data supporting this study are openly available from the University of Southampton repository (DOI: 10.5258/SOTON/XXXXXXXXX).

**Supplementary Materials:**

Materials and Methods

Figures S1-S8

Movies S1-S3

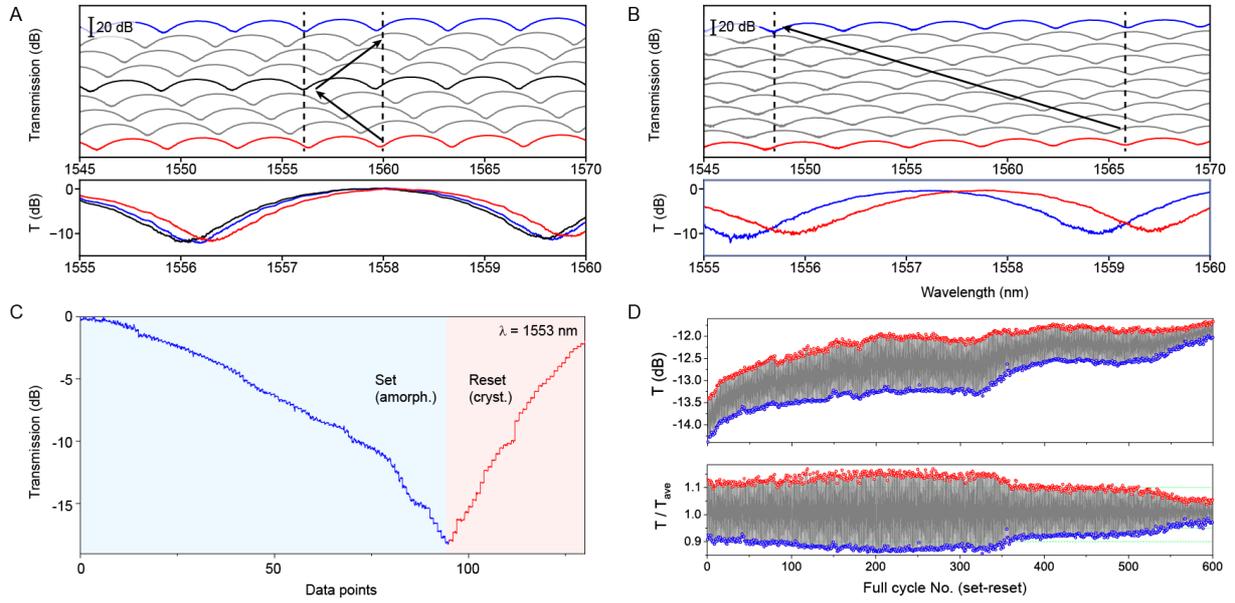

**Fig. 1. Tuning of a Mach-Zehnder interferometer by switching of an ultralow-loss PCM for selective optical phase control. A,B** Spectral response of an MZI during phase tuning using laser annealing of a cladding phase change material from the initial state (blue) though intermediate states (grey and black) to the final spectrum (red), vertically offset (top), and normalized to the initial state (bottom). A reversible $2\pi$ phase shift (**A**) and $10\pi$ tuning (**b**) with no change to modulation depth or insertion loss. **C** Transmission of an MZI for a fixed wavelength as a 50 µm patch is amorphized (blue) and recrystallized over 20 µm (red). **d**, The transmission though an MZI as a single spot is amorphized and recrystallized 600 times (top). The same data with the drift normalized using a 50-point moving average (bottom).
8

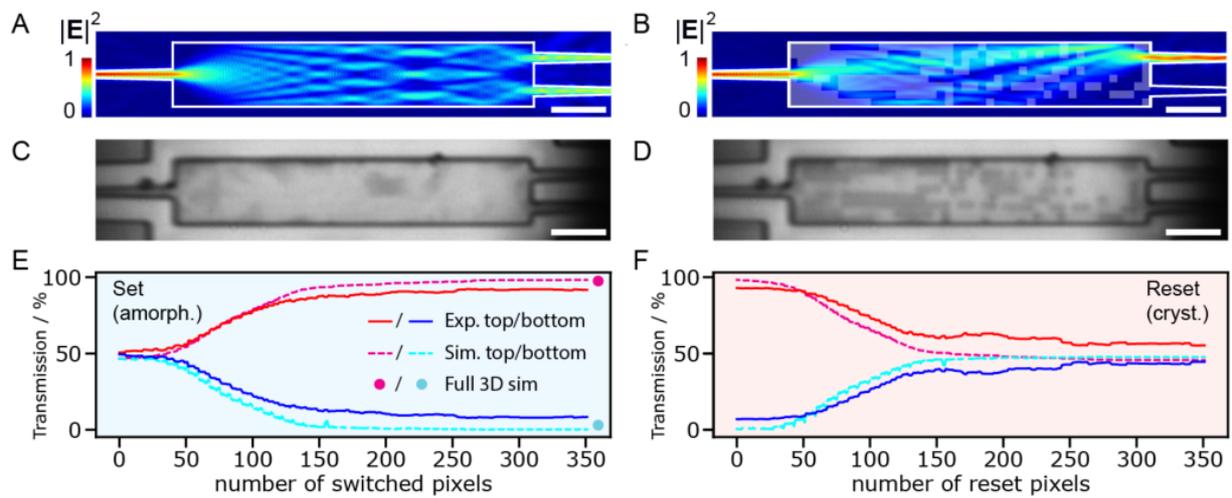

**Fig. 2 Programming of an MMI 1x2 splitter by writing a PCM pixel pattern. A,B** Simulated electric field distribution for an unperturbed (**A**) MMI clad with 23 nm of crystalline $Sb_2Se_3$ and with a perturbed MMI with the pixel pattern of amorphous $Sb_2Se_3$ overlayed in transparent white (**B**). **C,D** Optical images of an experimental MMI clad with unperturbed crystalline $Sb_2Se_3$ (**C**) and with an amorphous pixel pattern written into the $Sb_2Se_3$ (**D**). All scale bars, 5 µm. **E,F** Simulated transmission using 2D optimization model (green and teal dashed lines) for the top and bottom outputs of the MMI as a function of pixels written (set, amorphization) (**E**) and recrystallized (reset) (**F**) compared to the experimental values (red and blue solid lines). Dots represent final values obtained from 3D simulations with optimized pattern.

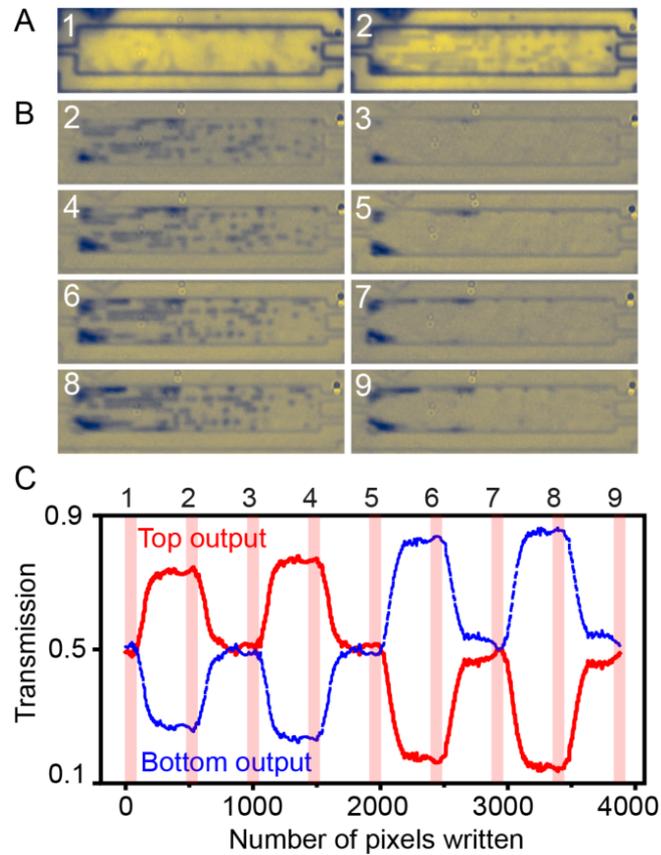

**Fig. 3 Reconfigurable optical routing between two outputs within an MMI.** **A** Optical images of an MMI in the original state (1) and with first pixel pattern set (2). **B** Processed images of set (2,4,6,8) and reset (3,5,7,9) states obtained by differencing with the original image (1). Set states are designed for routing to top output (2,4) and bottom output (6,8). **C** Continuous transmission ratio for the top (red) and bottom (blue) outputs of the MMI during the pattern writing and rewriting, numbers corresponding to maps in **A** and **B**.

# Supplementary Materials for

# Non-volatile programmable silicon photonics using an ultralow-loss $Sb_2Se_3$ phase change material

Matthew Delaney, Ioannis Zeimpekis, Han Du, Xingzhao Yan, Mehdi Banakar, David J. Thomson, Daniel W. Hewak, Otto L. Muskens

Correspondence to: O.Muskens@soton.ac.uk

**This PDF file includes:**

Materials and Methods
Supplementary Text
Figs. S1 to S8
Captions for Movies S1 to S3

**Other Supplementary Materials for this manuscript include the following:**

Movies S1 to S3

**Materials and Methods**

Materials

The devices shown in this work were fabricated using an industrial deep UV scanner on a 220 nm SOI platform with a 120 nm etch depth. The UV scanner was used to open windows above the MMI and MZI regions, and 23 nm thin $Sb_2Se_3$ films were deposited using low temperature RF sputtering. Following removal of the photoresist, the whole chip was then capped with a 200 nm thick layer of $ZnS:SiO_2$, using RF sputtering.

Methods

A 150 mW, 638 nm wavelength single mode diode laser was used to crystallize and amorphize the $Sb_2Se_3$, using pulses of 50 ms at 19 mW and 400 ns at 35 mW respectively. A 0.42 NA 50x objective was mounted to a 3D piezo stage, normal to the devices under test, which was for focusing the laser and taking optical images with a white light source and CCD camera. Using the piezo stage the objective was moved to create the pixel patterns with high reproducibility. Spectral data was taken with an Agilent 8163B swept source from 1530 to 1600 nm at a power of 10 mW. A modulated narrow linewidth laser at 1550 nm was used for transmission data, in combination with a lock-in amplifier.

The optical setup is shown in Fig. S1. Light from a telecommunications-band laser (Keysight) was sent through an optical fibre and was coupled into the device under test (D.U.T) using grating couplers. Transmitted light was collected by a second fibre and detected using an InGaAs photodetector (Thorlabs). The laser source was modulated at a kHz frequency, and a lock-in amplifier (Zurich Instruments) was used to extract the component from the photodetector signal. A custom-built two-fibre arm was used for measurements involving two outputs of the multimode interference devices (MMIs), using two detectors and two lock-in channels to simultaneously detect the individual transmitted intensities in each output of the MMI.

The second optical path consists of a directly modulated diode laser at 638 nm wavelength and with maximum power output of 170 mW in continuous wave operation. The light was linearly polarized and was transmitted with 100% efficiency through a polarizing beam splitter (PBS). A small fraction (<10%) of power is lost by coupling in a white-light source for inspection using a glass slide beamsplitter. A quarter waveplate was used between the PBS and the microscope to rotate the reflected light in order to route it to the inspection camera. A 50x, NA 0.55 microscope objective was used with working distance of 13mm to focus the write laser onto the D.U.T. The position of the objective is controlled using a closed-loop 3-axis piezo stage (Thorlabs) with 20 µm range in all three axes. In some of the experiments, an additional motorized translation stage was used for scanning in one direction along the waveguide.

A computer-controlled short pulse generator (Berkeley Nucleonics) was used to modulate the diode laser on time scales from nanoseconds to seconds. A home-built Labview program was used to control the experiment.

**Supplementary Text**

Insertion losses of MZI devices

Insertion loss of devices was measured for the Mach-Zehnder interferometers (MZIs) presented in Figure 2 of the main text. Figure S2 shows results on MZIs with patches of the PCM $Sb_2Se_3$ of different lengths from 0 µm (without $Sb_2Se_3$) – 750 µm deposited on the bottom arm of the asymmetric MZI (long arm). We observe that the addition of a 50 µm patch $Sb_2Se_3$ does not affect the optical response of the MZI. An oscillation in the maximum extinction over the spectral range is attributed to small (<1%) variations in the intensity balance between the two arms, possibly related to the MMI splitters used in the device. Between 50 µm and 125 µm patch length, this oscillation shifts and the maximum extinction is obtained around 1570nm wavelength. Importantly both 125 µm and 250 µm PCM-clad devices show good maximum, indicating that the addition of the PCM results in only small difference in the balance of the arms.

The result in Figure S2 can be interpreted by considering that the initial devices have slightly (around 1%) more intensity in the bottom arm in the spectral range 1560 nm – 1580 nm and therefore the MZI is unbalanced in this range (Figure S2a). At the same time the devices are balanced in the spectral ranges 1530 nm – 1550 nm and 1580nm – 1600 nm where the on-off extinction reaches >20dB. By adding patches of PCM of up to 250 µm length in the bottom arm, the attenuation in this arm results in an improved balance in the 1560 nm – 1580 nm range, while the devices become increasingly unbalanced toward the other two spectral bands with increasing patch length (Figure S2b-d). For even longer PCM patches >250 µm in length, the entire device is unbalanced as intensity in the PCM-cladded arm is further reduced. A variation of ± 1% of the splitting ratios of the MZI arms is attributed to the wavelength response of the MMI splitters.

Figure S3 shows the spectrally averaged insertion loss (IL) of the MZIs obtained from Figure S2 by averaging the signal over the entire spectral bandwidth and subtracting the IL of the straight waveguide (black lines in Figure 2). The IL of the reference straight waveguide includes the grating couplers. The IL versus length of the PCM patch shows a flat value of between 2 – 3 dB which is characteristic of the bare device and includes the throughput of the two MMI splitters.

Endurance testing of the PCM material
Figure S4 shows an additional experiment where the endurance of the $Sb_2Se_3$ PCM was tested. Compared to the result presented in the main text Fig. 2D, this run showed a more stable device response over the entire phase-change cycling. We point out that the time between full switching cycles is around 2s, therefore every endurance test represents several hours in which the experiment needs to remain stable within 100 nm on the waveguide. For this run we report >500 full cycles (write-reset) without notable degradation.

Above 570 cycles, we observe a period of instability however the instrument settles back into a stable switching with only a few additional jumps. We note that the jumps are primarily observed in the amorphization (blue dots) while the crystallization (red dots) remains relatively stable. Both the drift observed in Fig. 2D of the main text and the fluctuations of this dataset indicate some temporary instability of local domains in amorphization which needs to be further optimized.

Control over individual pixels in complex patterns
Writing of two-dimensional patterns in the $Sb_2Se_3$ phase-change layer requires a very high level of control over individual pixels. Pixels have to be consistent in size and shape, and should be unperturbed by adjacent domains that have already been exposed. In our work we have achieved control over pixels with pitch size as small as 750 nm.

Figure S5 shows results for a test pattern written in the middle of an MMI device. Starting from the initial device (1) a single pixel was written (2), which was completed into a 2x2 array (3) with uniform dimensions and precise spacing between domains. Resetting of the bottom right pixel (4) resulted in full erasure without affecting adjacent domains.
Next, the pattern was expanded to a 3x3 array (5). Here, the centre pixel was successfully reset (6) and rewritten (7) again demonstrating the capabilities of the technique in controlling individual pixels in complex arrangements. Finally, the device was reset (8) showing a local morphology similar to the original device.

Iterative MMI optimization
In our studies we investigated the use of patterns generated by iterative optimization. For this purpose, for every pixel a test was performed to see whether switching increases the transmission at the output port. The initially crystalline MMI was switched to the amorphous phase while intensity at both output ports was monitored. If the intensity at the target output increased, the switch pixel was kept, while the pixel was reset when switching decreased the target output intensity. In this experiment, the entire device was optimized in one pass (pass 1), followed by a second pass which started again at the beginning of the device and which re-optimized the first 20 µm of the MMI (pass 2).

Figure S6a,b shows results for bottom (blue) and top (red) output ports where only the accepted pixels are shown, every data point corresponds to a switched or reset pixel in the MMI. Figure S6c,d shows the same data on a logarithmic scale and normalized to the original port intensities of the unperturbed device, $T_0$. Raw port output signals are subject to intrinsic variations in grating efficiency and collection fibres, therefore normalizing the signals looks at relative changes compared to the unperturbed device.

Figure S6b shows an increase of single port transmission of the bottom output by 2 dB, accompanied by a 6 dB reduction of the top output. The total transmission, defined as the sum of the unnormalized $T_{top} + T_{bottom}$, shows an overall 0.5 dB increase in IL caused by writing the pattern. The changes induced by writing the pixel pattern are reversible and the original device transmission is retrieved by resetting all pixels in the device as shown in Figure S6b,d,f. Movies of writing (both passes 1 and 2) and resetting the device are included in the Supplementary material. Figure S7 shows snapshots of the device after pass 1 (1), pass 2 (2), and after completely resetting the device (3).

A second example of an optimized MMI is given in Figure S8. In this device three subsequent write passes were done over the full device length. While qualitatively the same results are obtained as in Figure S6, a notable difference is the increase in total device transmission by patterning, indicating that writing of pixels resulted in a better throughput. We attribute this effect to some losses in the crystalline state and compensation of impedance mismatch of the original device by shaping of the wavefront.

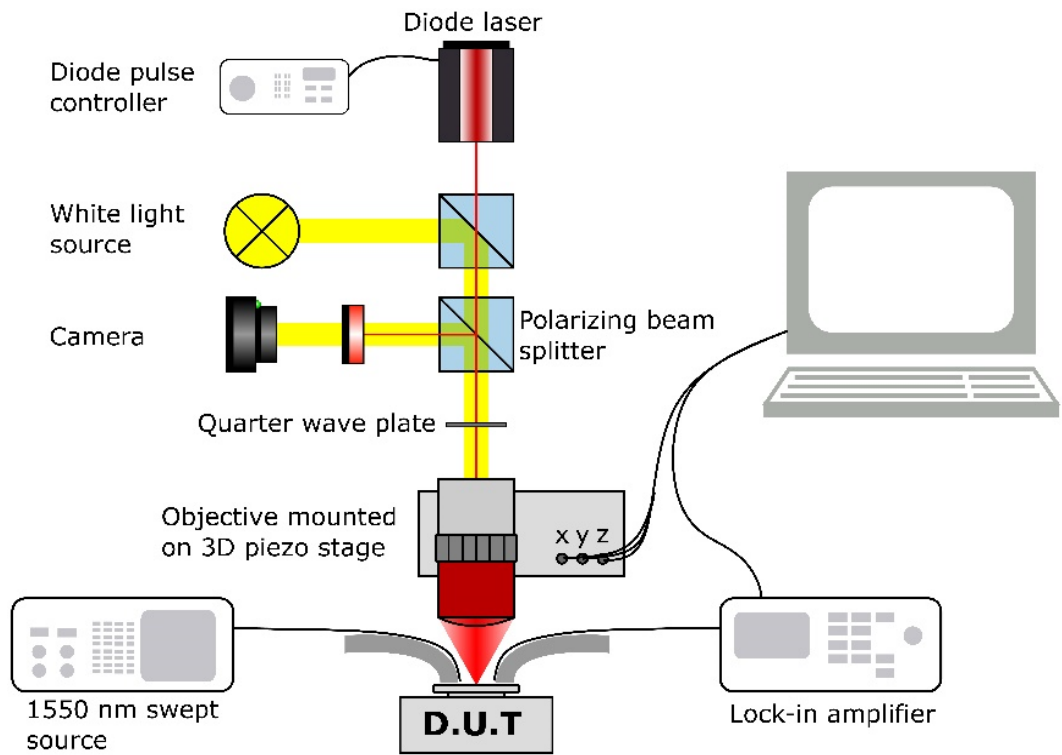

**Fig. S1.**
Optical setup for writing of pixels onto silicon photonic waveguide devices cladded with $Sb_2Se_3$.

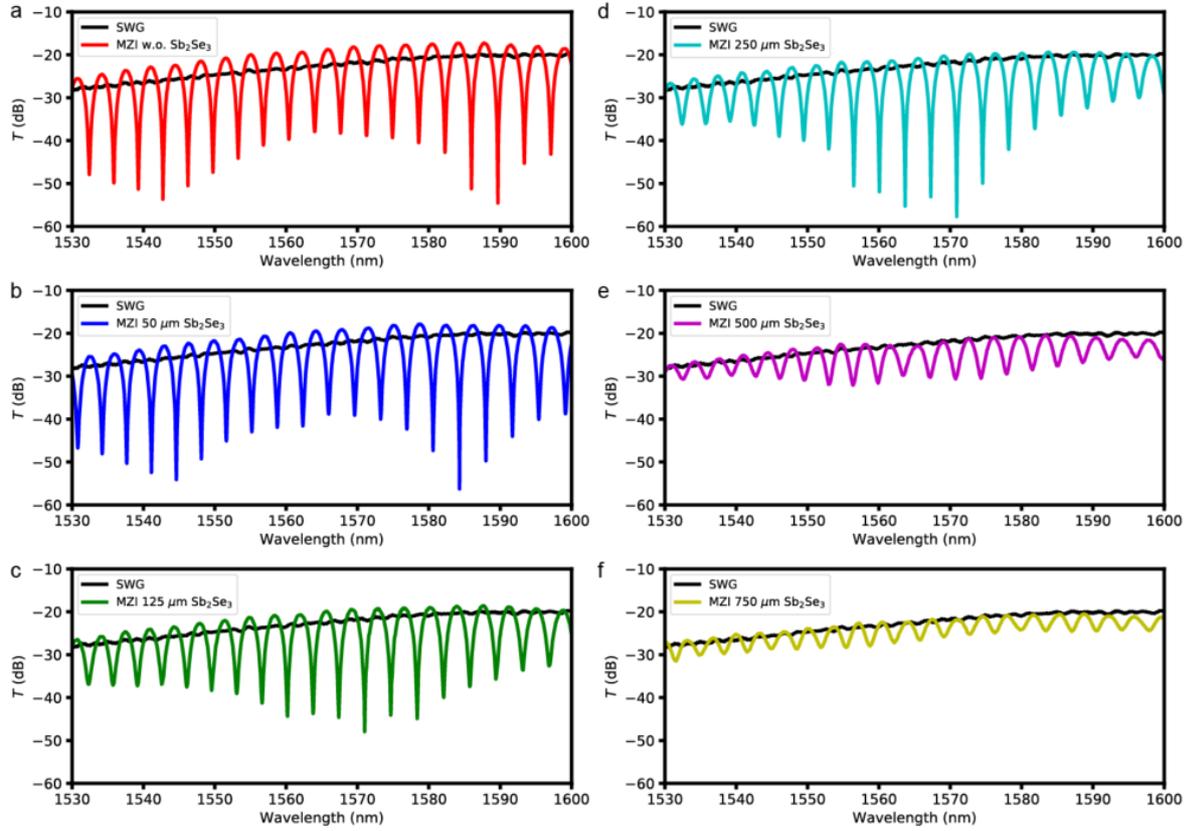

**Fig. S2.**

(A-F) Transmission spectra of control MZI without PCM (A) and MZIs with $Sb_2Se_3$ PCM patches of varying length (B-F). Results are compared to a straight waveguide (SWG) without PCM. For PCM lengths >250 µm, the MZI extinction contrast is reduced due to unbalancing of the arms by the PCM losses. All PCM patches were crystallized using a hot plate prior to measurements.

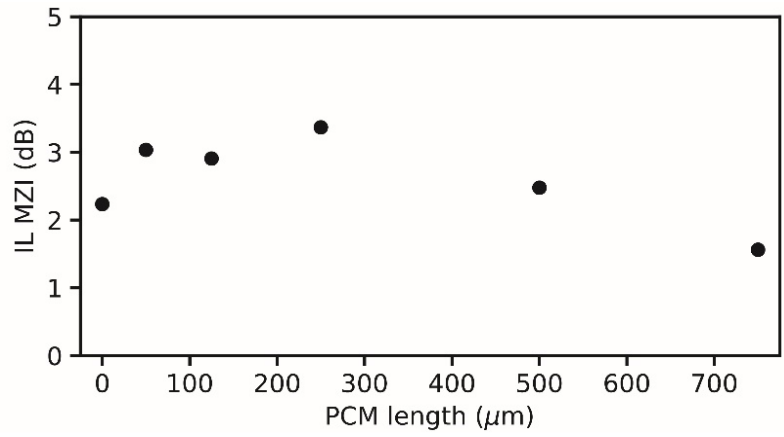

**Fig. S3.**
Insertion loss of the MZIs compared to SWG, for different lengths of the $Sb_2Se_3$ PCM patch.

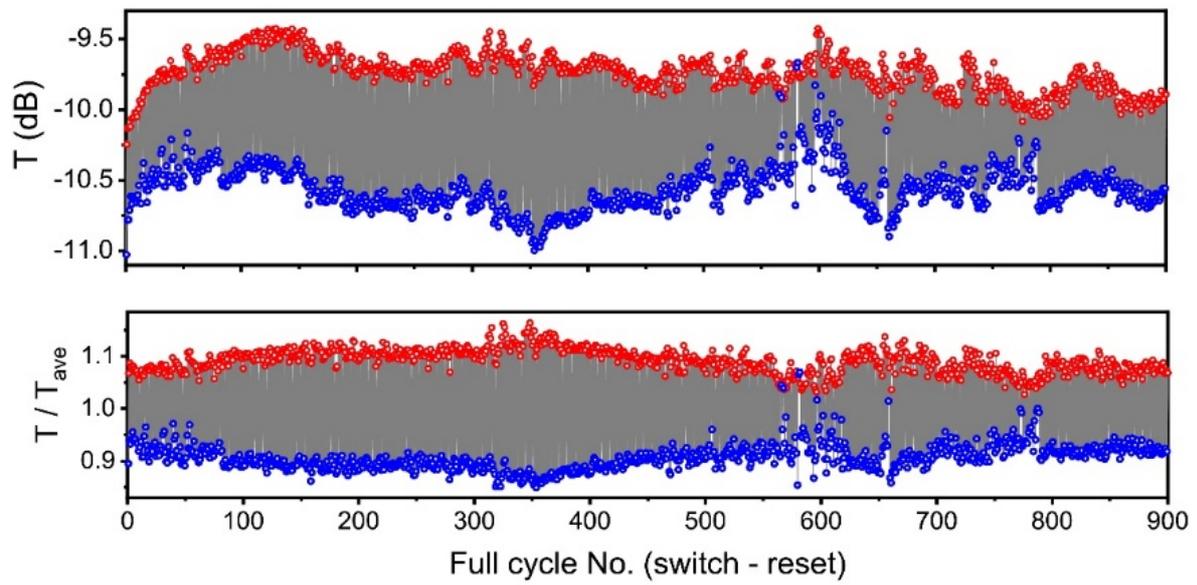

**Fig. S4.**
Example of endurance cycling experiment, showing transmission of MZI normalized to maximum against full switching cycle, each cycle consisting of a switch - reset pulse pair. Blue data: amorphization part, red data: crystallization part of the cycle.

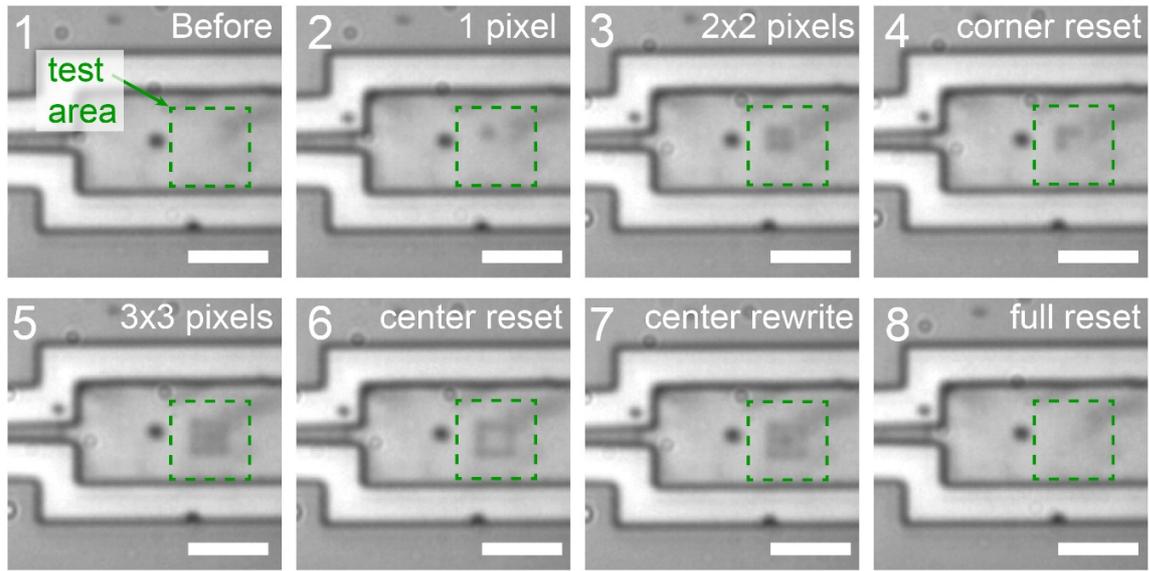

**Fig. S5.**
Demonstration of control over individual pixels in a patterned MMI. Subsequent pixel operations as shown in panels (1-8): device before writing (1), isolated single pixel writing (2), 2x2 array of pixels at 750 nm spacing (3), resetting of bottom right pixel (4), expansion to 3x3 array of pixels (5), resetting of central pixel in dense array (6), rewriting of the central pixel (7) and full reset (8). Scale bars, 5 µm.

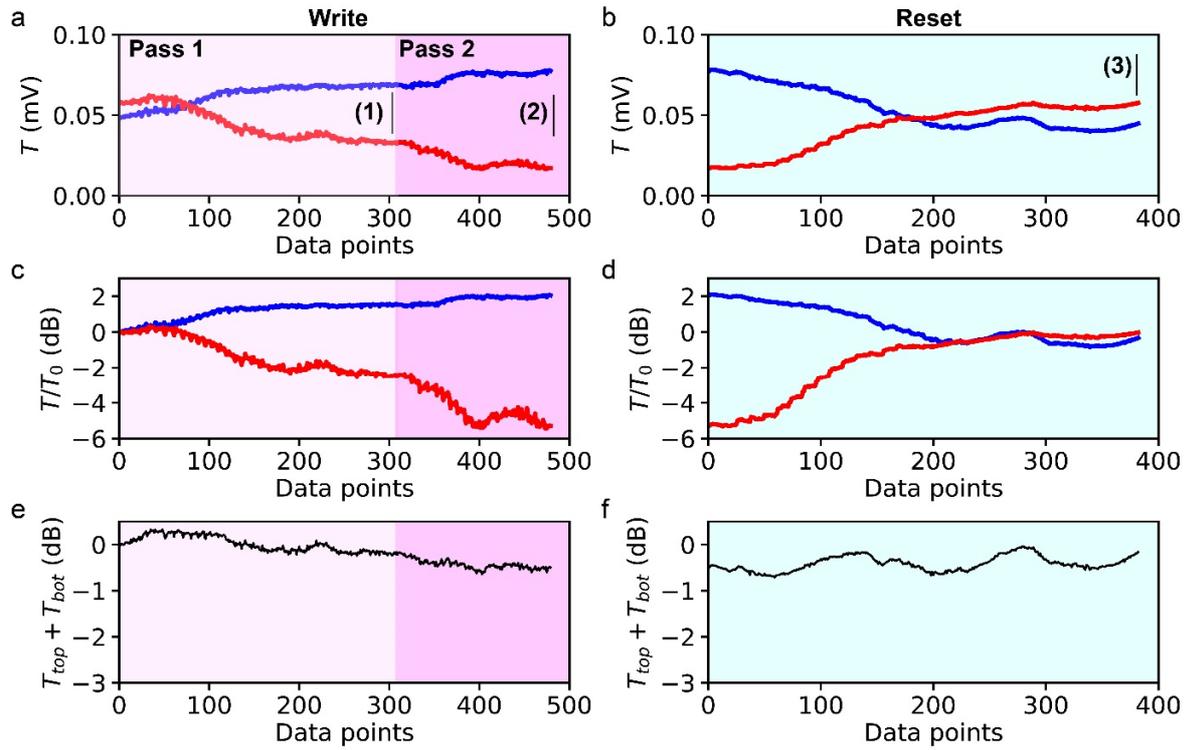

**Fig. S6.**
Demonstration of iterative optimization scheme showing intensities at top and bottom MMI outputs simultaneously detected using dual fibre output under conditions of iterative optimization (A,C,E) and device reset (B,D,F) at 750 nm pixel pitch. Only results for accepted pixels are shown. Panels represent raw signal intensities (A,B), log transmission normalized to the value of the unperturbed device (C,D), and the sum of top and bottom outputs normalized to the unperturbed device showing a 0.50 dB excess loss of the patterned device. Labels (1) – (3) refer to locations of snapshots of the final patterns of pass 1 and 2, as well as the fully reset device as presented in Fig. 7 of the main text.

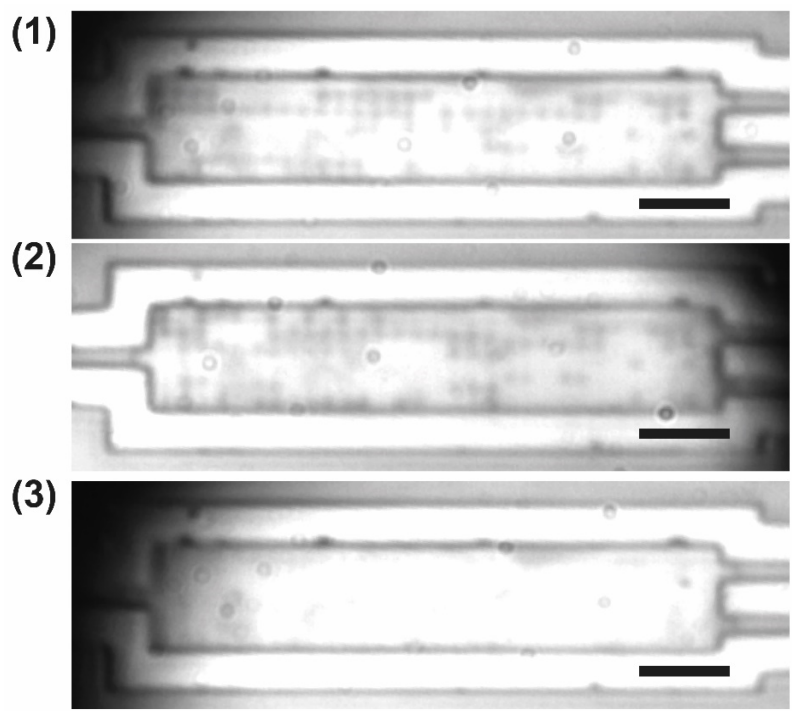

**Fig. S7.**
Images taken at various selected points in the writing process corresponding to labels in Fig. S6. A set of videos of the entire iterative writing process and the reset is available as Supplementary Materials Movies S1-S3. Scale bars, 5 µm.

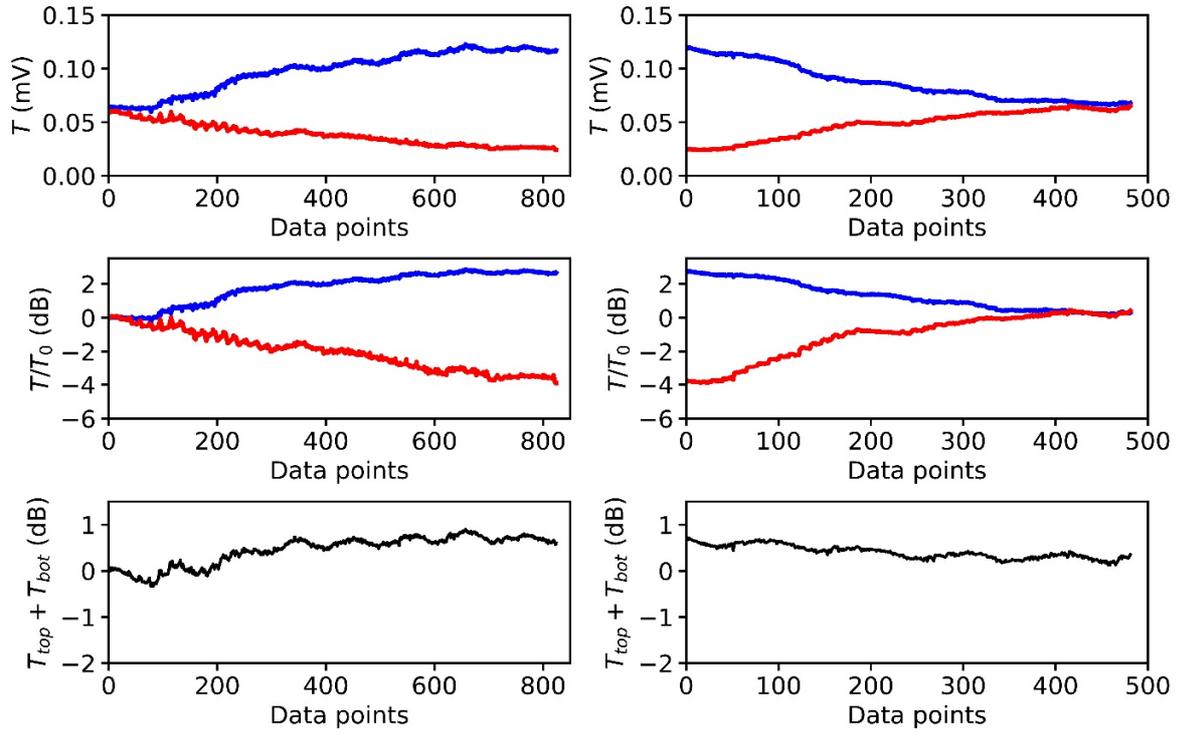

**Fig. S8.**
Another demonstration of iterative optimization scheme showing intensities at top and bottom MMI outputs simultaneously detected using dual fibre output under conditions of iterative optimization (A, C, E) and device reset (B, D, F) at 750 nm pixel pitch. Only results for accepted pixels are shown. Panels represent raw signal intensities (A, B), log transmission normalized to the value of the unperturbed device (C, D), and the sum of top and bottom outputs normalized to the unperturbed device showing a 0.60 dB total throughput increase of the patterned device compared to the unpatterned MMI. Three full cycles of optimization were done.